\documentclass[12pt]{article}

\usepackage{scicite}

\usepackage{graphicx}

\usepackage{times}

\newenvironment{sciabstract}{%
\begin{quote} \bf}
{\end{quote}}

\title{Spontaneously polarized half-gapped superconductivity} 

\author
{Sheng Ran,$^{1,2\ast}$ Chris Eckberg,$^2$ Qing-Ping Ding,$^{3}$ Yuji Furukawa,$^{3}$ Tristin Metz,$^2$\\ Shanta R. Saha,$^{1,2}$ I-Lin Liu,$^{1,2,4}$ Mark Zic,$^2$\\ Hyunsoo Kim,$^2$ Johnpierre Paglione,$^{1,2}$ Nicholas P. Butch$^{1,2\ast}$
\\
\normalsize{$^{1}$ NIST Center for Neutron Research,}\\
\normalsize{National Institute of Standards and Technology, Gaithersburg, MD 20899, USA}\\
\normalsize{$^{2}$ Center for Nanophysics and Advanced Materials, Department of Physics,}\\
\normalsize{University of Maryland, College Park, MD 20742, USA}\\
\normalsize{$^{3}$ Ames Laboratory, U.S. DOE and Department of Physics and Astronomy,}\\
\normalsize{Iowa State University, Ames, Iowa 50011, USA}\\
\normalsize{$^{4}$ Department of Materials Science and Engineering,}\\
\normalsize{University of Maryland, College Park, MD 20742 USA}\\
\\
\normalsize{$^\ast$To whom correspondence should be addressed; E-mail:  sran@umd.edu and nbutch@umd.edu.}
}

\date{}

\begin{document}

\baselineskip24pt

\maketitle 

\begin{sciabstract}
Nonunitary superconductivity is a rare and striking phenomenon in which spin up and spin down electrons segregate into two different quantum condensates. Because they support topological excitations, such superconductors are being seriously considered for potential quantum information applications. We report the discovery of nonunitary spin-triplet superconductivity in UTe$_{2}$, featuring the high transition temperature of 1.6~K and a remarkably large and anisotropic upper critical field exceeding 40~T. In this unusual superconducting state, electrons with parallel spins pair, yet only half of the available electrons participate, yielding a spin-polarized condensate that coexists with a spin-polarized metal. The superconducting order parameter, which breaks both gauge and time reversal symmetries, arises from strong ferromagnetic fluctuations, placing UTe$_{2}$ as the paramagnetic end member of the ferromagnetic superconductor series. This discovery yields a new platform for encoding information using topological excitations and for manipulation of spin-polarized currents.
\end{sciabstract}

One of the most interesting differences between spin triplet superconductors and the conventional spin singlet variety is the two-component triplet order parameter that allows spin up and down electrons to couple with different strength. Such nonunitary superconductors, in which spin up and down components have different gaps and an intrinsic spin polarization develops, are ideal platforms for studying topological phenomena~\cite{Kozii2016,Sau2012}. Nevertheless, nonunitary superconductors have received much less attention than their unitary equal-gapped counterparts, largely because they are rare to find. So far, the only established examples of nonunitary pairing include the superfluid $^3$He in high magnetic fields~\cite{Ambegaokar1973}, known as the A1 phase, as well as ferromagnetic superconductors~\cite{Machida2001}. It is an intriguing question whether nonunitary pairing can happen in the absence of a magnetic field - external or internal - thus spontaneously breaking time reversal symmetry.  

Here we report the discovery of novel nonunitary spin-triplet superconductivity in UTe$_{2}$, which closely resembles the ferromagnetic superconductors~\cite{Saxena2000,Aoki2001,Huy2007} with dramatically enhanced transition temperature and upper critical field, and a paramagnetic normal state. UTe$_{2}$ exhibits the crucial ingredients of a nonunitary triplet superconducting state, namely: an extremely large, anisotropic upper critical field {\itshape H}$_{c2}$, temperature independent nuclear magnetic resonance (NMR) Knight shift in the superconducting state that can only be due to triplet pairing, and a large residual normal electronic density of states indicating that half of the electrons remain ungapped. In other words, a spin up superfluid coexists with a spin down Fermi liquid. This discovery yields a new platform for encoding information using topological excitations and for manipulation of spin-polarized currents. 

UTe$_{2}$ crystallizes in the orthorhombic, centrosymmetric structure (space group 71 $Immm$). U atoms compose parallel linear chains oriented along the [100] {\itshape a}-axis (Fig.~1c), which coincides with the magnetic easy axis, as seen in the magnetic susceptibility $M/H$ (Fig.~2a). The low symmetry of this structure is responsible for the large magnetic anisotropy~\cite{Ikeda2006}, similar to the anisotropy in the orthorhombic, ferromagnetic superconductors URhGe and UCoGe~\cite{Aoki2001,Huy2007}. Unlike these compounds, or the isoelectronic compound USe$_{2}$~\cite{Noel1996}, the temperature dependence of the magnetization and electrical resistivity show no indications of a phase transition to a magnetically ordered state (Fig.~2). The high-temperature magnetization data show uncorrelated, paramagnetic behavior along all the three crystallographic axes. A Curie-Weiss fit yields an effective moment of 2.8 $\mu_B$/U, reduced from the value of a fully degenerate 5$f^2$ or 5$f^3$ configuration. At low temperatures, the magnetization increases sharply along the {\itshape a}-axis, while along the {\itshape b}-axis the magnetization decreases and becomes temperature-independent, a signature of Kondo coherence~\cite{Stewart1984}. 

The low-temperature magnetic behavior shows that UTe$_{2}$ is on the verge of ferromagnetism. Below 10 K, the {\itshape a}-axis magnetization exhibits neither conventional field/temperature (H/T) paramagnetic scaling, nor Arrott-Noakes ferromagnetic critical scaling~\cite{Butch2009} (see supplemental material Fig. 6). Instead, the data are well-described by the Belitz-Kirkpatrick-Vojta (BKV) theory of metallic ferromagnetic quantum criticality~\cite{Kirkpatrick2015}. For temperatures less than 9 K and fields less than 3 T, the magnetization data scale as $M/T^{\beta}$ vs. $H/T^{\beta + \gamma}$ (Fig.~2e), using BKV critical exponents ($\beta$ = 1, $\gamma$ = 0.5, $\delta$ = 1.5), behavior that has only otherwise been observed in NiCoCr$_{0.8}$ \cite{Sales2017}. This remarkable scaling, extending over five orders of magnitude, indicates that UTe$_{2}$ is a quantum critical ferromagnet, dominated by strong magnetic fluctuations. 

The high-temperature electrical resistivity $\rho(T)$ is typical of uncorrelated, paramagnetic moments in the presence of single-ion Kondo hybridization with the conduction band, which is responsible for the negative slope. At temperatures below a crossover marked by maximal resistivity, the Kondo hybridization yields coherent electronic bands, resulting in a metallic temperature-dependence (Fig.~2c, quantitative details in supplementary text). 

The transition from this correlated normal state to a superconducting ground state below the critical temperature {\itshape T}$_{c} = 1.6$~K is robust and sharp, evident in the low-temperature $\rho(T)$, ac magnetization and specific heat $C(T)$ data (Fig.~3). The normalized jump in $C(T)$ at {\itshape T}$_{c}$ is $\Delta C$/$\gamma T_{c}$ = 1.28 is comparable to the conventional Bardeen-Cooper-Schrieffer (BCS) value of 1.43 expected from weak coupling. However, there is a large residual value of the Sommerfeld coefficient $\gamma_0 = 55$~mJ/mol-K$^2$ in the superconducting state, from which it is immediately apparent that half of the electronic states in this material are not gapped by the superconducting transition, indicative of an unconventional pairing mechanism, as occurs in UPt$_{3}$, UCoGe, and UGe$_{2}$~\cite{Joynt2002,Mineev2017a}. Crucially, there is little variation in the residual $\gamma_0$ value between samples of UTe$_{2}$ with slightly different {\itshape T}$_{c}$ (Supplementary Material Fig. S11), implying that the large residual electronic density of states is an intrinsic, disorder-insensitive property of UTe$_{2}$ arising from nonunitary superconductivity. For temperatures below {\itshape T}$_{c}$, $C(T)$ follows a power law, with $n\sim$ 3.2, reflecting the presence of point nodes, which arise from a momentum-dependent gap structure typical of nonunitary states. 

Perhaps the most dramatic sign of unconventional superconductivity is obvious in the upper critical field {\itshape H}$_{c2}$ of this superconductor, which can only be explained by triplet pairing. The resistivity as a function of temperature for different magnetic fields applied along the three principle crystal axes is shown in Fig.~4. The {\itshape H}$_{c2}$ is strongly anisotropic, with the value along {\itshape b} exceeding the two orthogonal directions by a factor of 4 at 1~K. Incredibly, the zero temperature limit of {\itshape H}$_{c2}$ along {\itshape b} well exceeds the highest measured magnetic field of 20~T, and we conservatively estimate a value of 40~T based on the curvature of the critical field of that of UCoGe.

The upper critical field of a conventional singlet superconductor is restricted by both of the orbital and paramagnetic pair-breaking effects. The orbital limit in superconductors is often well described by the Werthamer-Helfand-Hohenberg (WHH) theory {\itshape H}$_{orb}$ = 0.7$dH_{c2} \over dT_{c}$$\mid_{Tc}T_{c}$~\cite{Helfand1966}. Although it can account for the response to field along the {\itshape a} axis, the WHH model is otherwise inapplicable, most prominently along the {\itshape b}-axis, where the slope of {\itshape H}$_{c2}$ at {\itshape T}$_{c}$ is about 17~T/K along {\itshape b}, which leads to an expected {\itshape H}$_{orb}$ = 20~T for this direction, dramatically disagreeing with the experimental curve. The conventional paramagnetic limit is given by {\itshape H}$_{para}$ = 1.86{\itshape T}$_{c}$ ~\cite{Clogston1962}, yielding {\itshape H}$_{para}$ = 3~T for UTe$_{2}$. It can be clearly seen that the experimental {\itshape H}$_{c2}$ value well exceeds {\itshape H}$_{para}$ in all three directions, quite remarkably by almost an order of magnitude along the {\itshape b} axis, excluding spin singlet order parameters. 

The violation of the orbital limit in directions perpendicular to the magnetic easy axis (the {\itshape a}-axis) is consistent with the behavior of the ferromagnetic superconductors~\cite{Mineev2017} and differs qualitatively from the relatively low {\itshape H}$_{c2}$ values found in other paramagnetic triplet superconductors~\cite{Mackenzie2003,Weng2016}. The unusual shape of the {\itshape H}$_{c2}$ curve of UTe$_2$ resembles those of UCoGe~\cite{Aoki2009} and URhGe~\cite{Levy2005}, in which ferromagnetic spin fluctuations are believed to mediate the superconducting pairs~\cite{Mineev2017a}. Even though the normal state of UTe$_{2}$ is not magnetically ordered, the striking similarities suggest that its superconducting pairs are also mediated by ferromagnetic spin fluctuations, placing it as the end member of the series of ferromagnetic superconductors. When superconducting pairing is mediated by ferromagnetic spin fluctuations, the field dependence of the magnetization is coupled to the field dependence of the superconducting coupling strength~\cite{Mineev2011}, as verified in UCoGe and URhGe~\cite{Wu2017}. The coupling strength $\lambda$ as a function of magnetic field can be estimated based upon the behavior of {\itshape H}$_{c2}$ and $\gamma$ (Ref. 24). Especially striking is the large increase in $\lambda$ along the {\itshape b}-axis of about 50$\%$ (Supplementary Material Fig. S5), which far exceeds the field-induced enhancement of $\lambda$ in UCoGe~\cite{Wu2017}. 

Further confirmation of spin triplet pairing in UTe$_{2}$ comes from NMR measurements, which are sensitive to internal magnetic fields (Fig~3d). No change of the peak position is observed in the $^{125}$Te-NMR spectra between normal and superconducting states, leading to a temperature-independent value of the $^{125}$Te Knight shift ($K$), which is proportional to the spin susceptibility of the quasiparticles forming the superconducting pairs. In singlet-paired superconductors, $K$ decreases below $T_c$, whereas in UTe$_2$, $K$ remains constant on passing through $T_c$, signifying that the superconducting pair is a spin triplet~\cite{Tou1996,Ishida1998}.

The simplest possible superconducting pairing symmetry consistent with all measured properties of UTe$_{2}$ is the nonunitary triplet state, in which a two-component superconducting order parameter has two different energy gaps. The experimental observation that half of the electronic states remain ungapped places UTe$_{2}$ in the extreme limit of nonunitary pairing, where one pairing state is completely suppressed and a spin up superfluid coexists with a spin down Fermi liquid. The superconducting wavefunction in UTe$_{2}$ is thus likely similar to the one derived for the ferromagnetic uranium superconductors~\cite{Mineev2017}.

Why does the superconductivity in UTe$_{2}$ gap one spin subband instead of both, as might be expected for a paramagnetic triplet superconductor~\cite{Hillier2012}? Generally, there is coupling between the magnetization and the superconducting triplet order parameter in the free energy~\cite{Hillier2012}. This coupling ensures that nonunitary triplet pairing is energetically favored in paramagnetic superconductors with a large magnetic susceptibility, which is especially enhanced near a ferromagnetic quantum critical point, precisely the condition found in UTe$_{2}$. Nonunitary order parameters generally break time reversal symmetry. We conclude that although the phase transition is driven by the critical fluctuations of the superconducting order parameter, the strong magnetic interactions in UTe$_{2}$ force an additional symmetry reduction, namely, the breaking of both gauge and time reversal symmetries in the superconducting state.

The discovery of this nonunitary superconducting state opens the door to advances in the study of topological electronic states and their application to quantum information technology. As a paramagnetic triplet superconductor, this material intrinsically consists of equal spin pairs, which may be useful for spin current generation~\cite{Linder2015}. Nonunitary superconductors also host Majorana excitations that may be detected by ARPES or STM~\cite{Sau2012}. The high {\itshape T}$_{c}$ and simple synthesis of this material makes this a very promising platform for making novel devices in which quantum spin states can be manipulated.

\bibliography{UTe2arxiv}

\bibliographystyle{Science}

\section*{Acknowledgments}
We acknowledge helpful discussions with Wesley Fuhrman, Yi-Ting Hsu, Tai Kong, Srinivas Raghu, Jay Sau, Yan Wang, Sheng-Long Xu, and Victor Yakovenko. \textbf{Funding:} Research at the University of Maryland was supported by the the National Science Foundation Division of Materials Research Award No. DMR-1610349, and the Gordon and Betty Moore Foundation’s EPiQS Initiative through Grant No. GBMF4419. Research at Ames Laboratory was supported by the U.S. Department of Energy (DOE), Office of Basic Energy Sciences, Division of Materials Sciences and Engineering. Ames Laboratory is operated for the U.S. DOE by Iowa State University under Contract No. DE-AC02-07CH11358. \textbf{Author contributions:} S. Ran and N. Butch conceived and designed the study. S. Ran synthesized the single crystalline samples. S. Ran, C. Eckberg, and I. Liu performed the electrical resistivity measurements. S. Ran, C. Eckberg, and T. Metz performed the specific heat measurements. S. Ran, S. R. Saha, and M. Zic performed the magnetization measurements. S. Ran and N. Butch performed the neutron scattering measurements. Q. Ding and Y. Furukawa performed the NMR measurements. All authors contributed to the preparation of the manuscript. \textbf{Competing Interests:} The authors declare no competing interests. \textbf{Data and materials availability:} All data are available in the manuscript or the supplementary materials.

\section*{Supplementary materials}
Materials and Methods\\
Supplementary Text\\
Figs. 5 to 18\\

\clearpage

\begin{figure}
\includegraphics[angle=0,width=160mm]{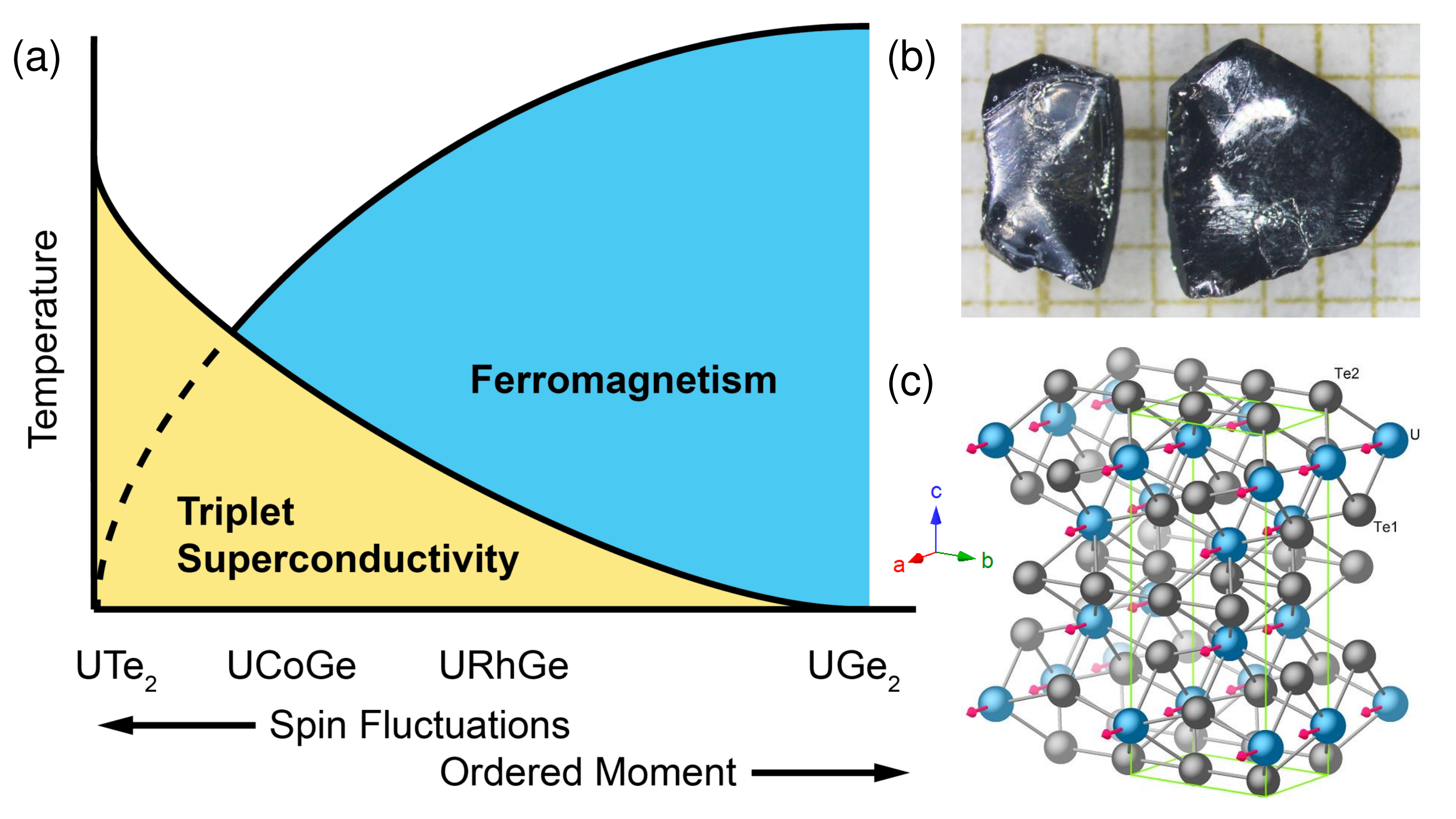} 
\caption{End member of the line of ferromagnetic superconductors. (a) Global phase diagram of ferromagnetic superconductors; UTe$_{2}$ is located at the paramagnetic end of the series. (b) The single crystals of UTe$_{2}$ grown using chemical vapor transport method on the millimeter scale. (c) Crystal structure of UTe$_{2}$, with U atoms in blue and Te atoms in gray. The U atoms sit on chains parallel to the [100] a-axis, which coincides with the magnetic easy axis, illustrated by the magenta arrows.} 
\label{cartoon}
\end{figure}

\begin{figure}
\includegraphics[angle=0,width=175mm]{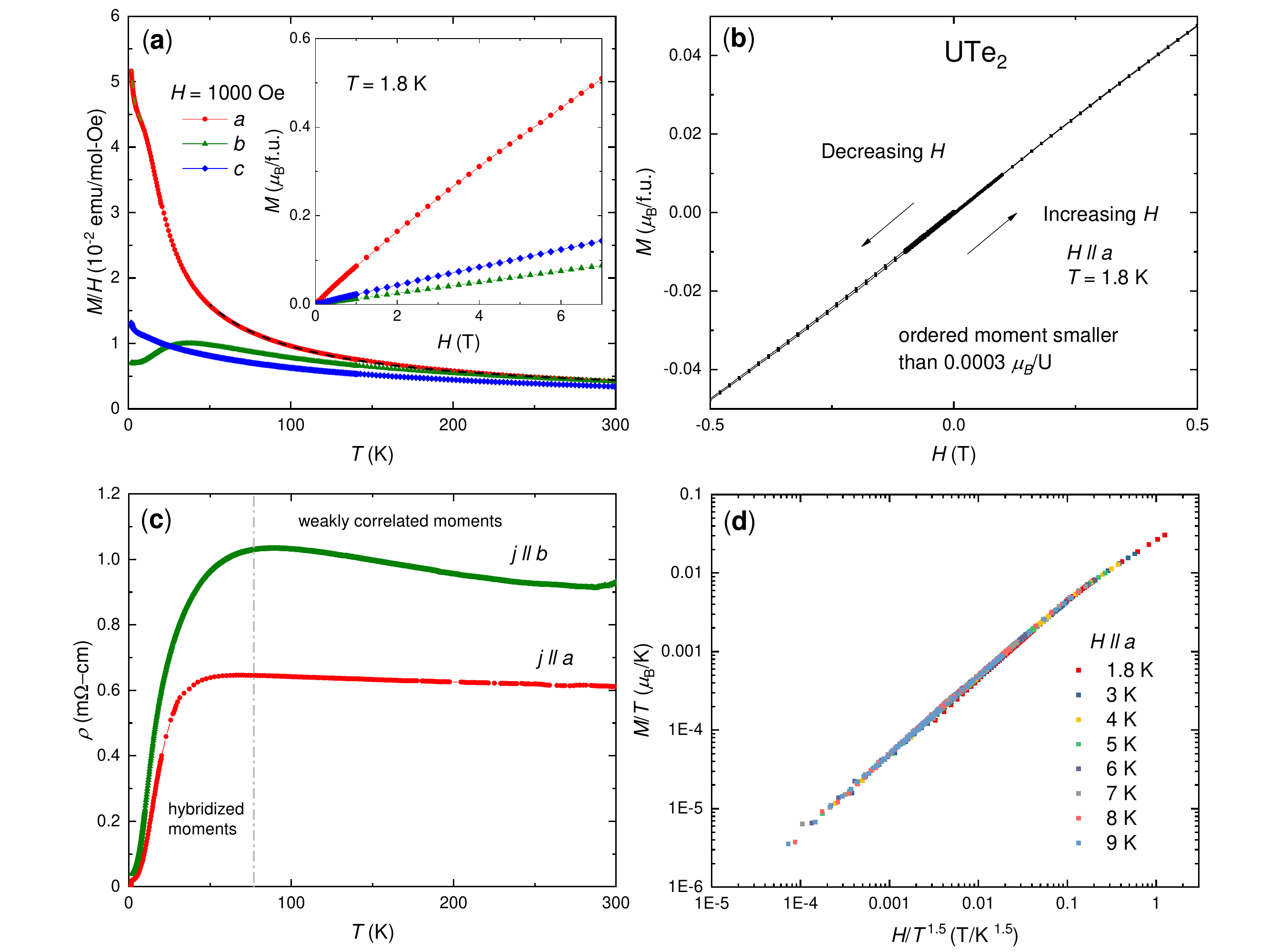}
\caption{Normal state properties of UTe$_{2}$. (a) Temperature dependence of magnetization with magnetic field of 0.1~T applied in three directions. The brown dashed line is the fit to the power law in the low temperature region, while the black dashed line is the fit to the Curie-Weiss law in the high temperature region. The inset shows the magnetization as a function of applied field in three directions at 2~K. (b) Magnetization data at 1.8~K upon increasing and decreasing magnetic field in the low field range showing no hysteresis. The upper limit for ordered moment is 0.0003 $\mu_B$/U obtained from the zero field magnetization value. (c) Temperature dependence of electronic resistivity data in zero magnetic field with electric current applied along {\itshape a} and {\itshape b}-axis. (d) $M/T$ as a function of $H/T^{1.5}$ for different temperatures. All the data collapse onto a single line. This behavior can be well described by BKV theory of metallic ferromagnetic QCP (see text).}
\label{NS}
\end{figure}

\begin{figure}
\includegraphics[angle=0,width=165mm]{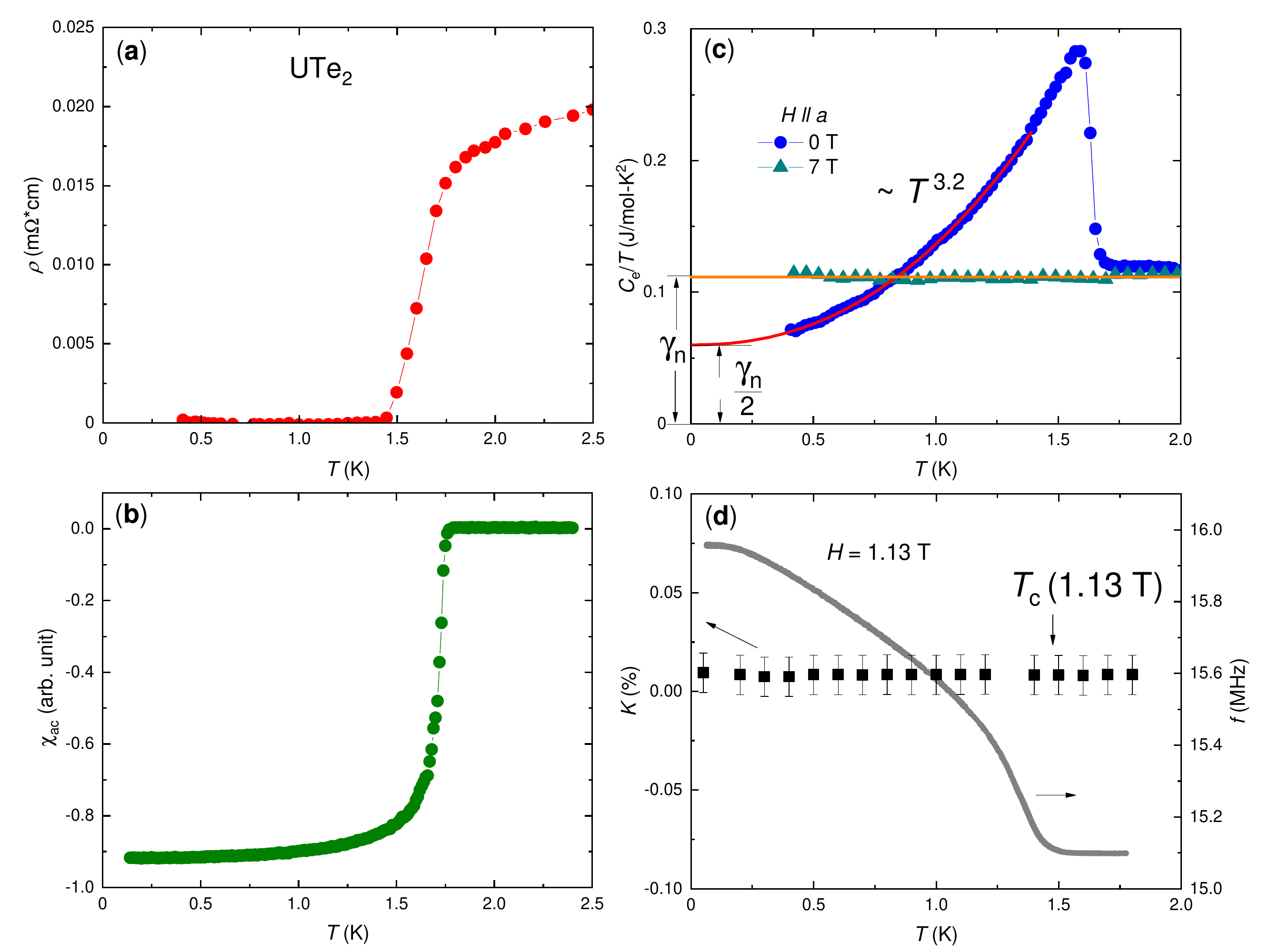} 
\caption{Superconducting state properties of UTe$_{2}$. Temperature dependence of (a) resistivity and (b) ac magnetization data at low temperatures showing the bulk superconductivity. (c) Electronic contribution to heat capacity (phonon contribution has been subtracted) in zero and 7~T, divided by temperature as a function of temperature to illustrate $\gamma$ in the superconducting and normal states. Magnetic field is applied along {\itshape a}-axis. (d) Temperature dependence of $^{125}$Te NMR Knight shift $K$ below and near $T_c$ of powdered UTe$_{2}$ sample (left axis) and temperature dependence of the resonance frequency $f$ of the NMR tank circuit confirming the superconducting state and $T_c$ (right axis).}
\label{SC}
\end{figure}

\begin{figure}
\includegraphics[angle=0,width=168mm]{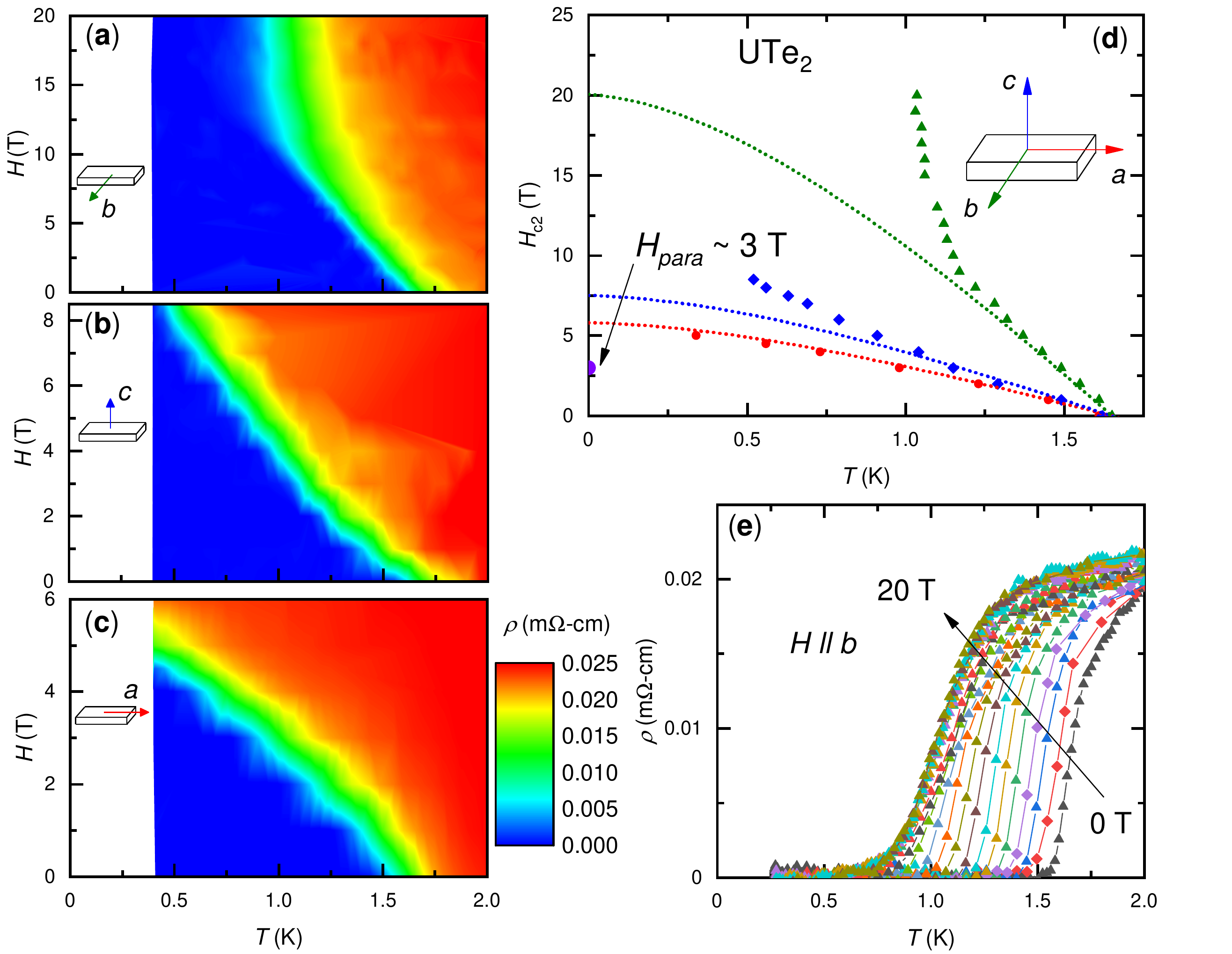} 
\caption{Upper critical field {\itshape H}$_{c2}$ of UTe$_{2}$. (a)-(c) Color contour plots of resistivity value as a function of temperature and magnetic field, with magnetic fields applied along (a) {\itshape b}-axis, (b) {\itshape c}-axis and (c) {\itshape a}-axis. The current is applied along {\itshape a}-axis. (d) The {\itshape H}$_{c2}$ value as a function of {\itshape T}$_{c}$ in three directions. Dotted lines represent the WHH fit of the {\itshape H}$_{c2}$ data. (e) Temperature dependent resistivity data in magnetic fields applied along $b$ axis up to 20 T.}
\label{Hc2}
\end{figure}

\clearpage

\section*{Materials and Methods}
Single crystals of UTe$_{2}$ were synthesized by the chemical vapor transport method using iodine as the transport agent. Elements of U and Te with atomic ratio 2:3 were sealed in an evacuated quartz tube, together with 3~mg/cm$^3$ iodine. The ampoule was gradually heated up and hold in the temperature gradient of 1060/1000~$^{\circ}$C for 7 days, after which it was furnace cooled to the room temperature. The crystal structure was determined by $x$-ray powder diffraction using a Rigaku $x$-ray diffractometer with Cu-K$_{\alpha}$ radiation. Crystal orientation was determined by Laue $x$-ray diffraction performed with a Photonic Science $x$-ray measurement system. Neutron scattering was performed on the NG-4 Disk Chopper Spectrometer at the NIST Center for Neutron Research. Electrical resistivity measurements were performed in a Quantum Design Physical Property Measurement System (PPMS) using the $^3$He option, and in Oxford $^3$He system. Magnetization measurements were performed in a magnetic field of 0.1~T using a Quantum Design Magnetic Property Measurement System (MPMS). AC magnetic susceptibility measurements were performed in a Quantum Design PPMS using the ADR option. Specific heat measurements were also performed in a Quantum Design PPMS using the $^3$He option, and in Oxford dilution refrigerator system. Ultra-low temperature NMR measurements of of $^{125}$Te ($I$ = 1/2, $\gamma_N$/2$\pi$ = 13.454~MHz/T) nuclei were conducted on powdered crystals using a lab-built phase coherent spin-echo pulse spectrometer and an Oxford dilution refrigerator installed at the Ames Lab. The $^{125}$Te-NMR spectra were obtained by sweeping the magnetic field $H$ at $f$ = 15.1~MHz. The data that support the results presented in this paper and other findings of this study are available from the corresponding authors upon reasonable request. Identification of commercial equipment does not imply recommendation or endorsement by NIST.

\section*{Supplementary Text}
\subsection*{X-ray and neutron diffraction}
Room temperature powder x-ray diffraction on crushed single crystals shows that CVT-grown UTe$_{2}$ forms in the correct crystal structure and is single phase, with no sign of impurity phases. Low-temperature neutron diffraction confirms that there are no structural or magnetic phase transitions down to 5~K.

\subsection*{Electrical resistivity}
The low temperature resistivity can be fit to Fermi liquid term $AT^2$ (Fig.~\ref{Rfit}), with $A\sim$ 0.64~$\mu$$\Omega$-cm/K$^2$ for {\itshape a}-axis and 1.55~$\mu$$\Omega$-cm/K$^2$ for {\itshape b}-axis. Values of RRR range from 18 to 30. These do not exhibit a large variation across different batches of single crystals synthesized via CVT.

The Kondo-coherent state exhibits strongly-renormalized Fermi liquid properties: 1) resistivity $\rho$ = $AT^2$, with $A\sim$ 1~$\mu$$\Omega$-cm/K$^2$, 2) specific heat C = $\gamma T$ with $\gamma$ = 120 mJ/mol-K$^2$, and 3) the Kadowaki - Woods ratio $A/\gamma^2$ $\sim$ 1$\times$10$^{-4}$~$\mu$$\Omega$-cm/K$^2$/(mJ/mol-K$^2$)$^2$, similar to many heavy fermion metals. 

\subsection*{Magnetization}
 
The Arrott plots (Fig.~\ref{Arrottplot}) in the low field range (0 - 0.1~T) at different temperatures show that the system is not in the critical regime of a mean-field classical (finite-temperature) ferromagnetic phase transition. Extending this analysis beyond mean field using the Arrott-Noakes equation of state is also unsuccessful.

The magnetization data are well-described by the Belitz-Kirkpatrick-Vojta (BKV) theory of metallic ferromagnetic quantum critical point. To determine critical exponents, the low temperature magnetization data was fitted to power law behavior, with $\gamma$ = 0.51 (Fig.~\ref{MTfit}). The consideration of a constant susceptibility $M/H$, consistent with a large Pauli paramagnetic response from the heavy Fermi liquid, is necessary to conform to established theories of ferromagnetic critical behavior. A constant term in $M/H$, or equivalently a linear term in $M(H)$, is subtracted from the measured $M(H)$ data. After the subtraction, for temperatures less than 9 K and fields less than 3 T, the resultant curves collapse onto a single curve when $M/T^{\beta}$ is plotted vs. $H/T^{\beta + \gamma}$ (main text, Fig.~2), using BKV critical exponents ($\beta$ = 1, $\gamma$ = 0.5, $\delta$ = 1.5). Note that scaling is also possible absent this correction. Without constant $M/H$ subtraction, $M/T^{\beta}$ vs. $H/T^{\beta + \gamma}$ data can also collapse onto a single curve, for temperatures less than 9 K and fields less than 7 T (Fig.~\ref{Mscaling2}). However the corresponding exponents will be $\beta$ = 4.16, $\gamma$ = 0.51, $\delta$ = 1.12. The small value of $\delta$ reflects the almost-linear $M(H)$, but the very large value of $\beta$ cannot be reconciled with known theories.

\subsection*{Specific heat}

The low-temperature $T^{3}$ phonon contribution to the specific heat is estimated by fitting to linear function to $C/T$ vs $T^{2}$ (Fig.~\ref{Cphonon}). It can also be seen that there are no signatures of magnetic phase transitions or unusual temperature-dependence above the superconducting $T_c$.

The deviation from BCS behavior of the superconducting transition in UTe$_{2}$ is emphasized in Fig.~\ref{Cfit}, in which it is clear that exponential temperature dependence expected for an isotropic gap is absent in this material. Instead, the specific heat below {\itshape T}$_{c}$ follows a power law, with $n\sim$ 3.2, reflecting the presence of point nodes, which arise from a momentum-dependent gap structure typical of nonunitary states. 

The large residual $\gamma$ is a robust feature and does not show obvious sample variation as seen in Fig.~\ref{Cvariation}. This fact is in sharp contrast to the strong sample dependence observed in other materials considered to house spin-triplet superconductivity.

$C/T$ data in the magnetic fields applied along $a$-axis are shown in Fig.~\ref{Cfield}. The residual $\gamma$ increases systematically upon increasing magnetic field, further indicating this is an intrinsic property of the compound, as magnetic field will enhance spin unbalance. Entropy calculated from specific heat data for superconducting and normal state are shown in Fig.~\ref{entropy}. The normal state data are obtained by applying a magnetic field of 7~T along the $a$-axis to suppress superconductivity. The superconducting jump releases 10$\%$ more entropy than expected, which can be ascribed to magnetic excitations arising from the spin-polarized ungapped normal Fermi liquid.

\subsection*{NMR}
No change of the peak position has been observed in the $^{125}$Te-NMR spectra between normal and superconducting states, as shown in Fig.~\ref{NMRSM}.

\clearpage

\begin{figure}
\centering
\includegraphics[angle=0,width=100mm]{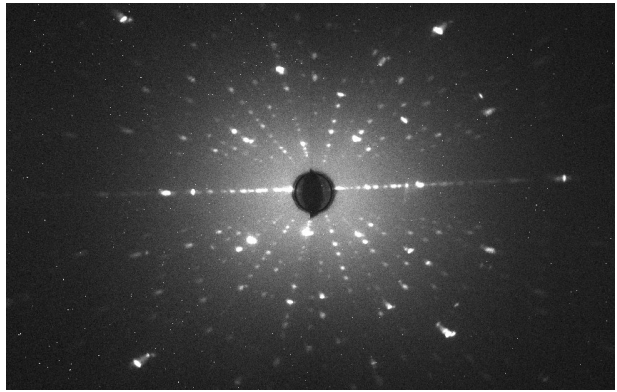}
\caption{Laue diffraction pattern of [011] direction demonstrating good crystallanity.}
\label{Laue}
\end{figure}

\begin{figure}
\centering
\includegraphics[angle=0,width=150mm]{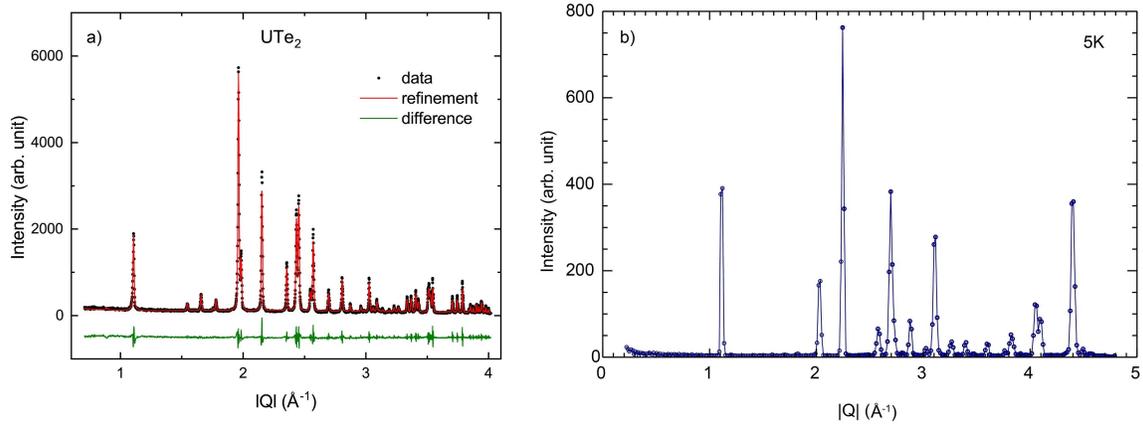}
\caption{(a) Powder $x$ray diffraction data of UTe$_{2}$ showing good quality of the sample with no visible peaks from impurities. (b) Low-temperature neutron diffraction data of UTe$_{2}$ confirming that there are no structural or magnetic phase transitions down to 5~K.}
\label{diffraction}
\end{figure}

\begin{figure}
\centering
\includegraphics[angle=0,width=100mm]{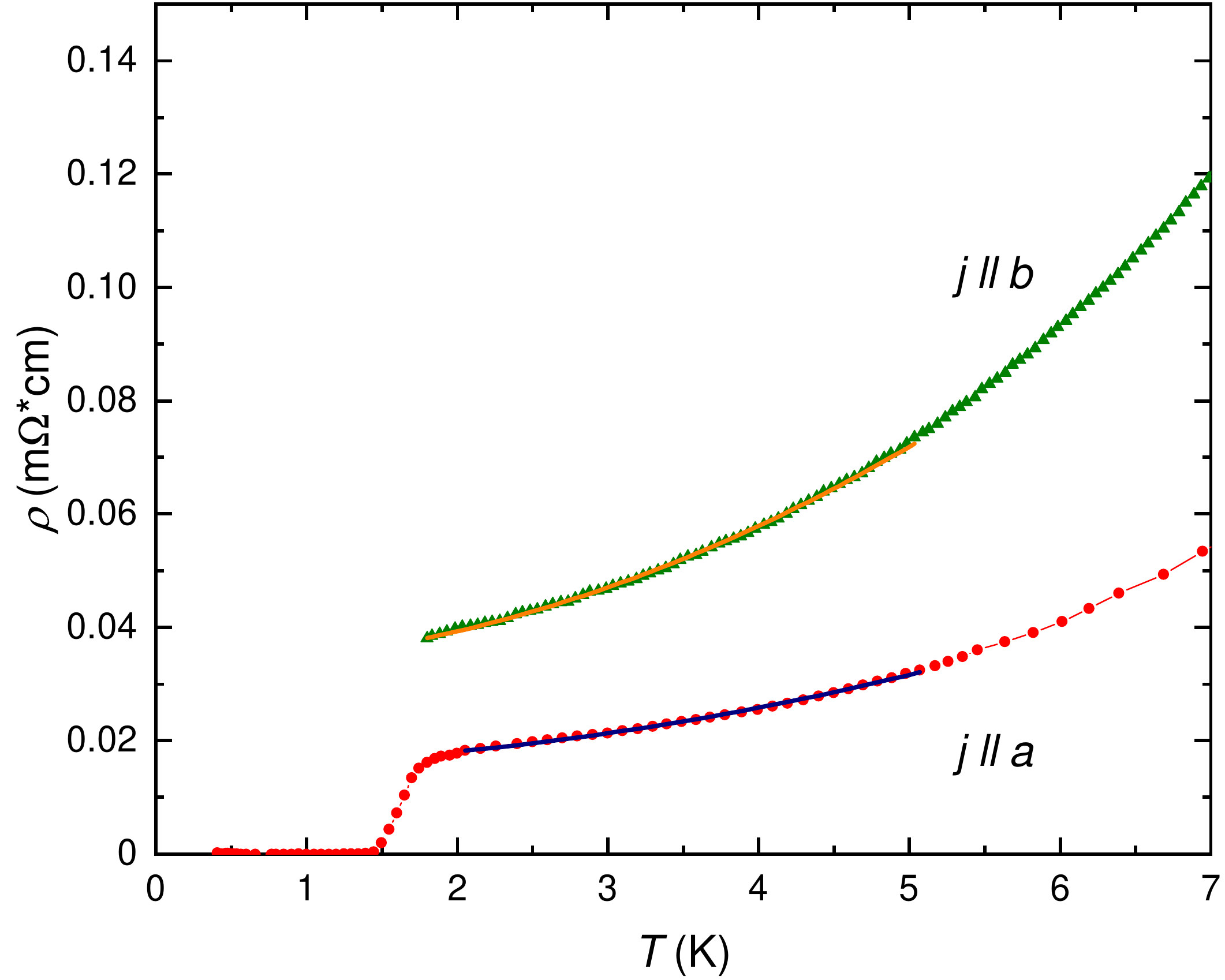}
\caption{Temperature dependence of electronic resistivity data in zero magnetic field with electric current applied along {\itshape a} and {\itshape b}-axis. The lines are the fit to Fermi liquid term $AT^2$.}
\label{Rfit}
\end{figure}

\begin{figure}
\centering
\includegraphics[angle=0,width=150mm]{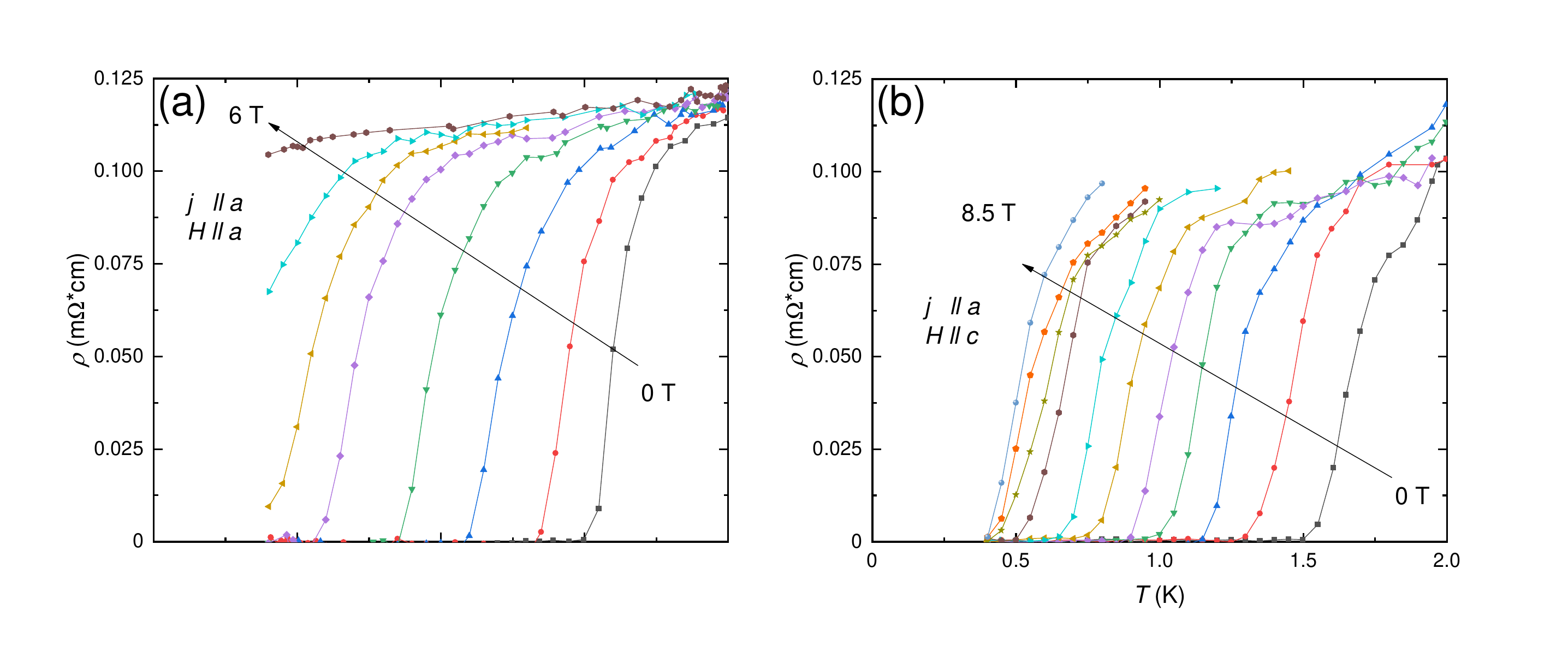}
\caption{Temperature dependent resistivity data in magnetic fields applied along (a) $a$ and (b) $c$ axis up to 9 T. The current is applied along {\itshape a}-axis.}
\label{Rfit}
\end{figure}

\begin{figure}
\centering
\includegraphics[angle=0,width=100mm]{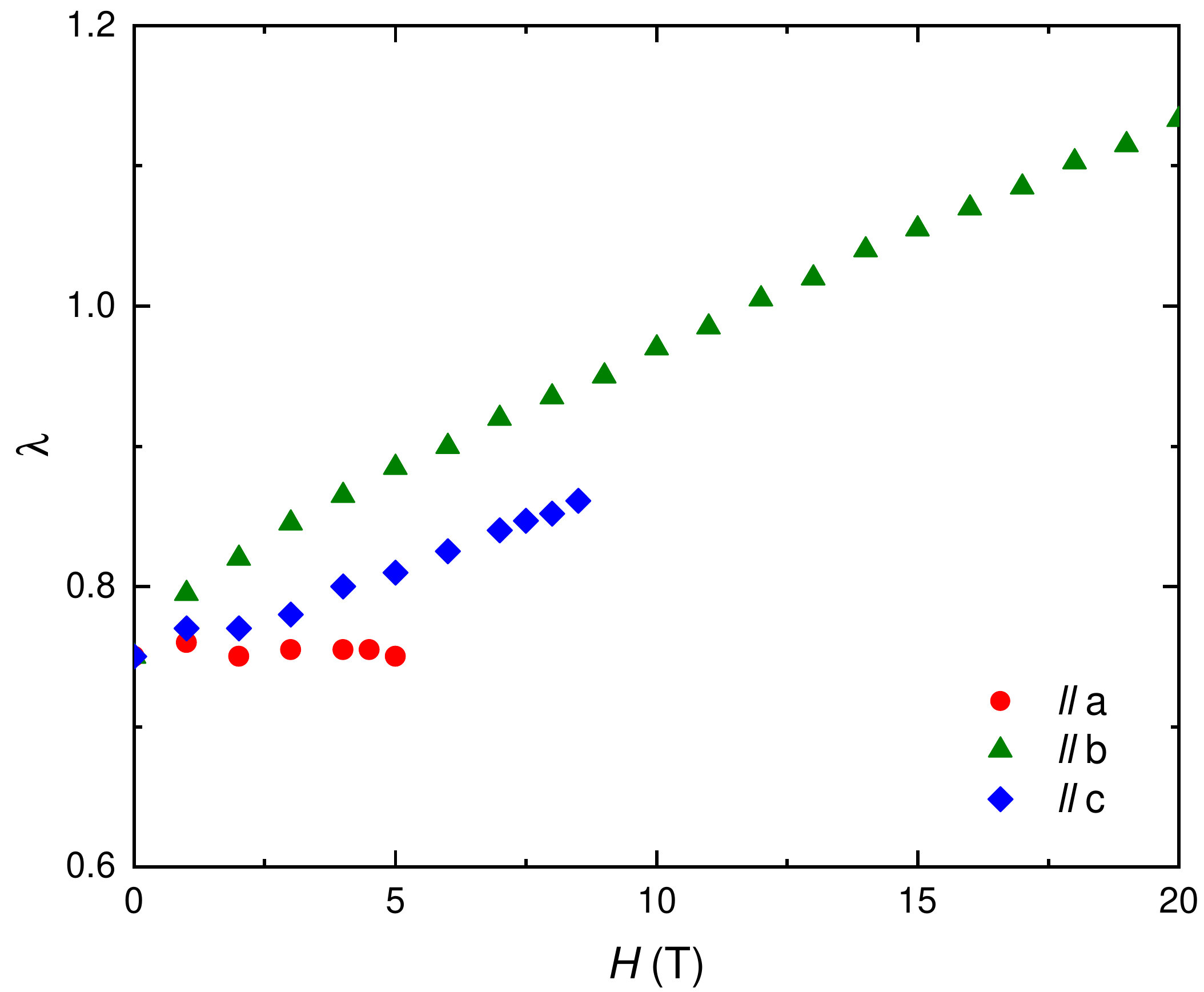}
\caption{The calculated superconducting coupling strength as a function of applied magnetic field in three directions is enhanced when field is applied along the {\itshape b}-axis, as expected from pairing due to ferromagnetic fluctuations.}
\label{coupling}
\end{figure}

\begin{figure}
\centering
\includegraphics[angle=0,width=100mm]{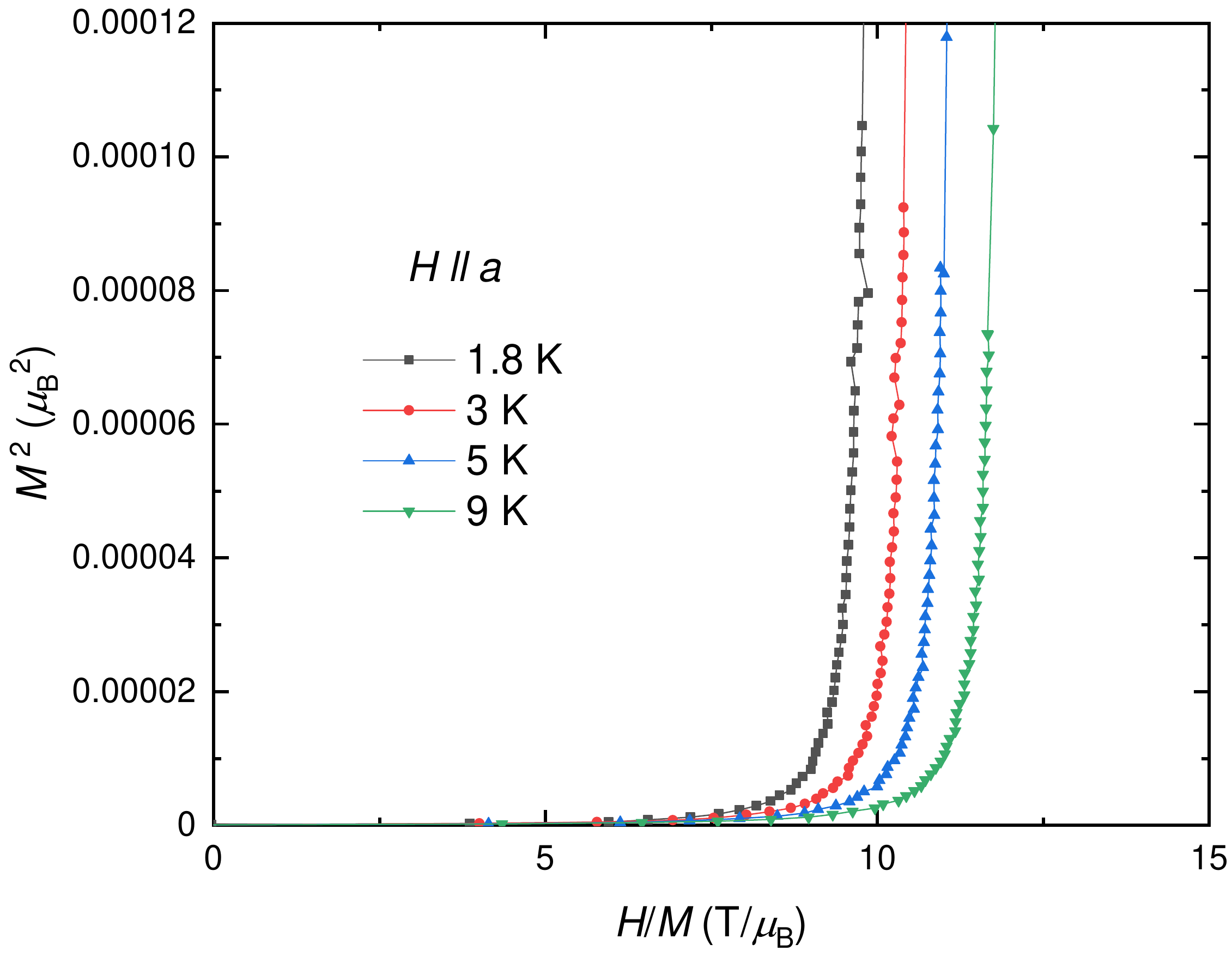}
\caption{Arrott plot, $M^2$ as a function of $H/M$, at different temperatures above $T_c$. It can be seen that UTe$_{2}$ does not have a conventional finite-temperature ferromagnetic transition.}
\label{Arrottplot}
\end{figure}

\begin{figure}
\centering
\includegraphics[angle=0,width=100mm]{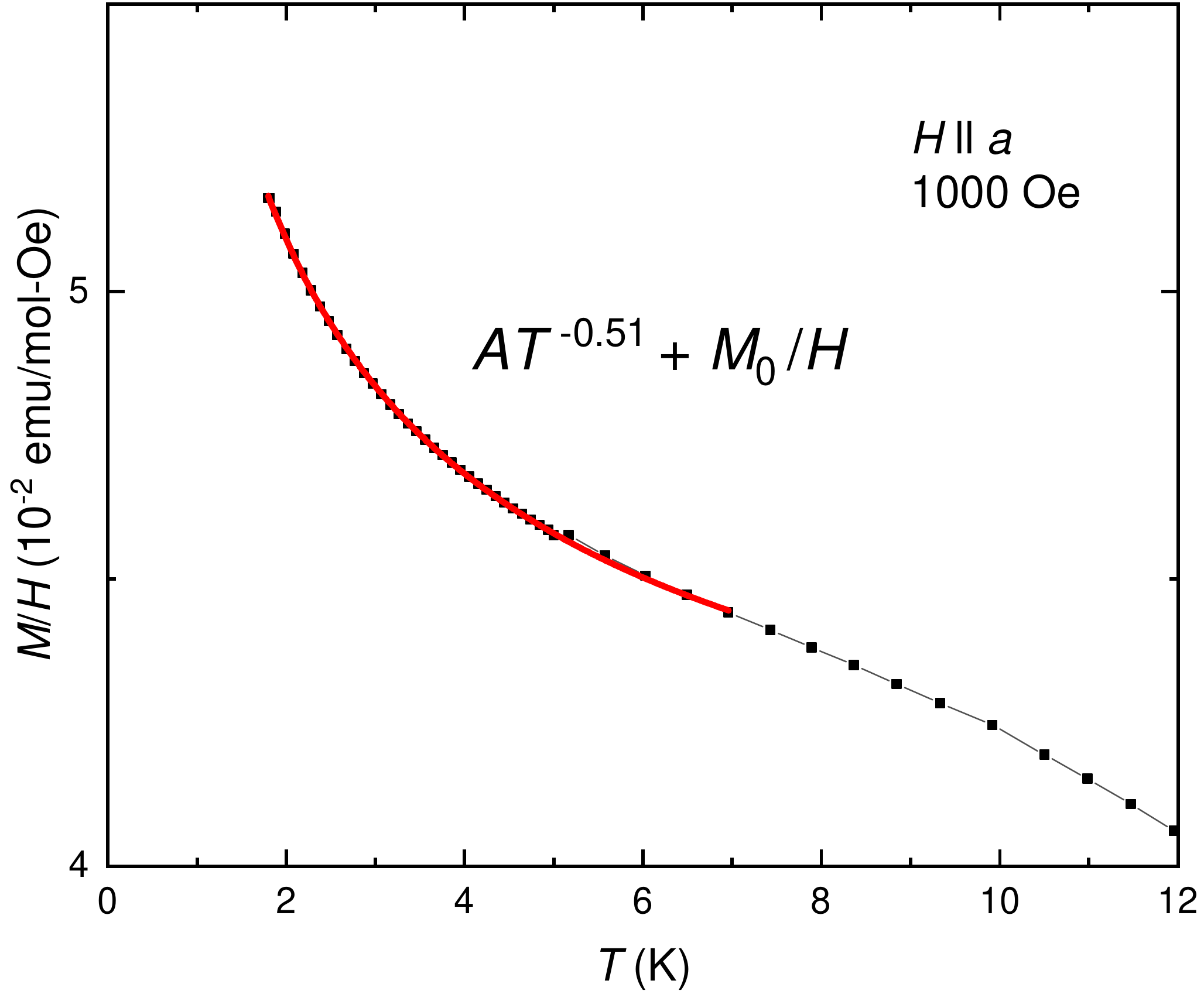}
\caption{Temperature dependence of magnetization with magnetic field of 0.1~T applied in $a$ axis. The red line is the fit to the power law $AT^{\upsilon} + M_0/H$ in the low temperature region. The constant term $M_0$ is necessary to obtain a good fitting.}
\label{MTfit}
\end{figure}

\begin{figure}
\centering
\includegraphics[angle=0,width=100mm]{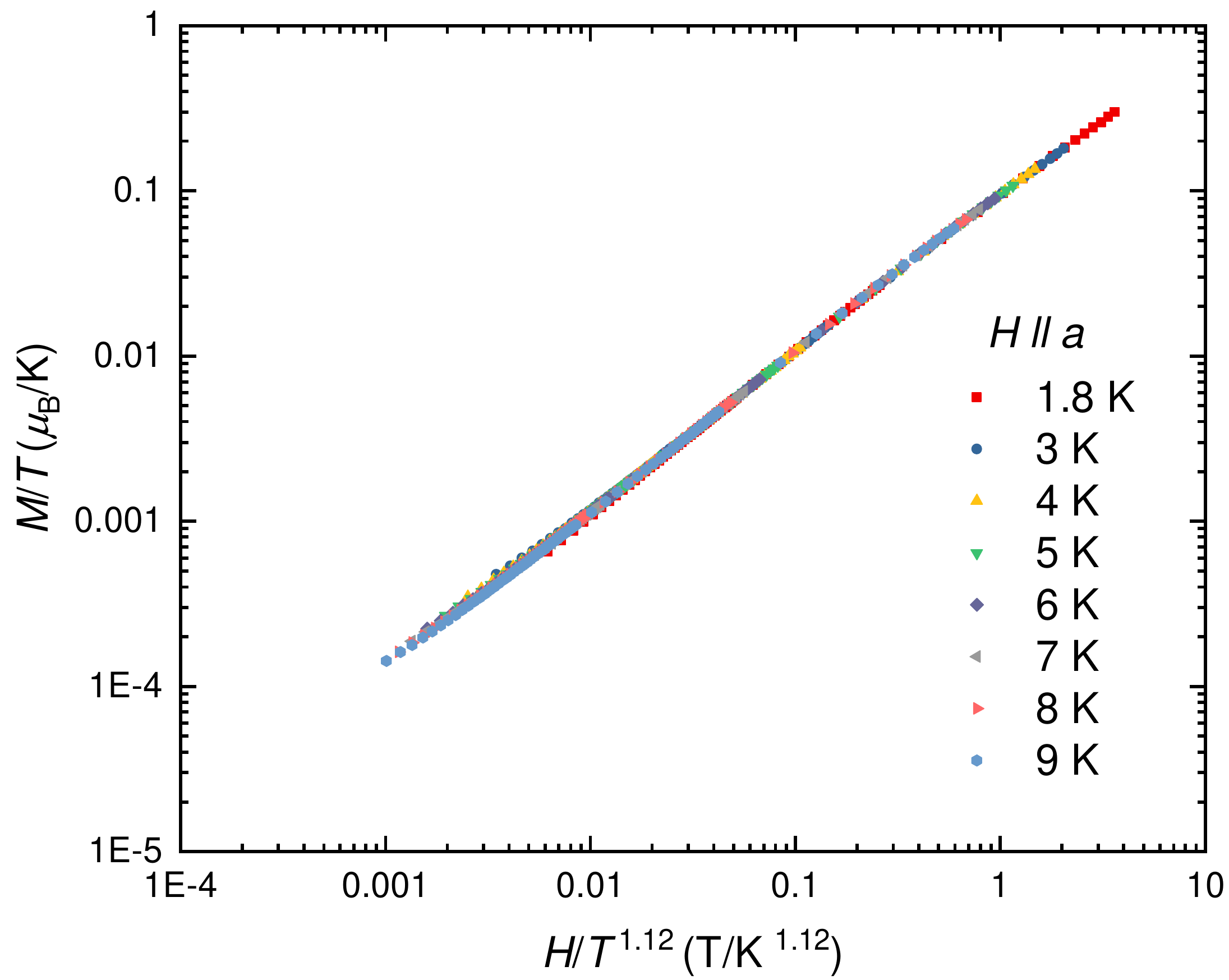}
\caption{$M/T$ as a function of $H/T^{1.12}$ for different temperatures. All the data collapse onto a single line.}
\label{Mscaling2}
\end{figure}

\begin{figure}
\centering
\includegraphics[angle=0,width=100mm]{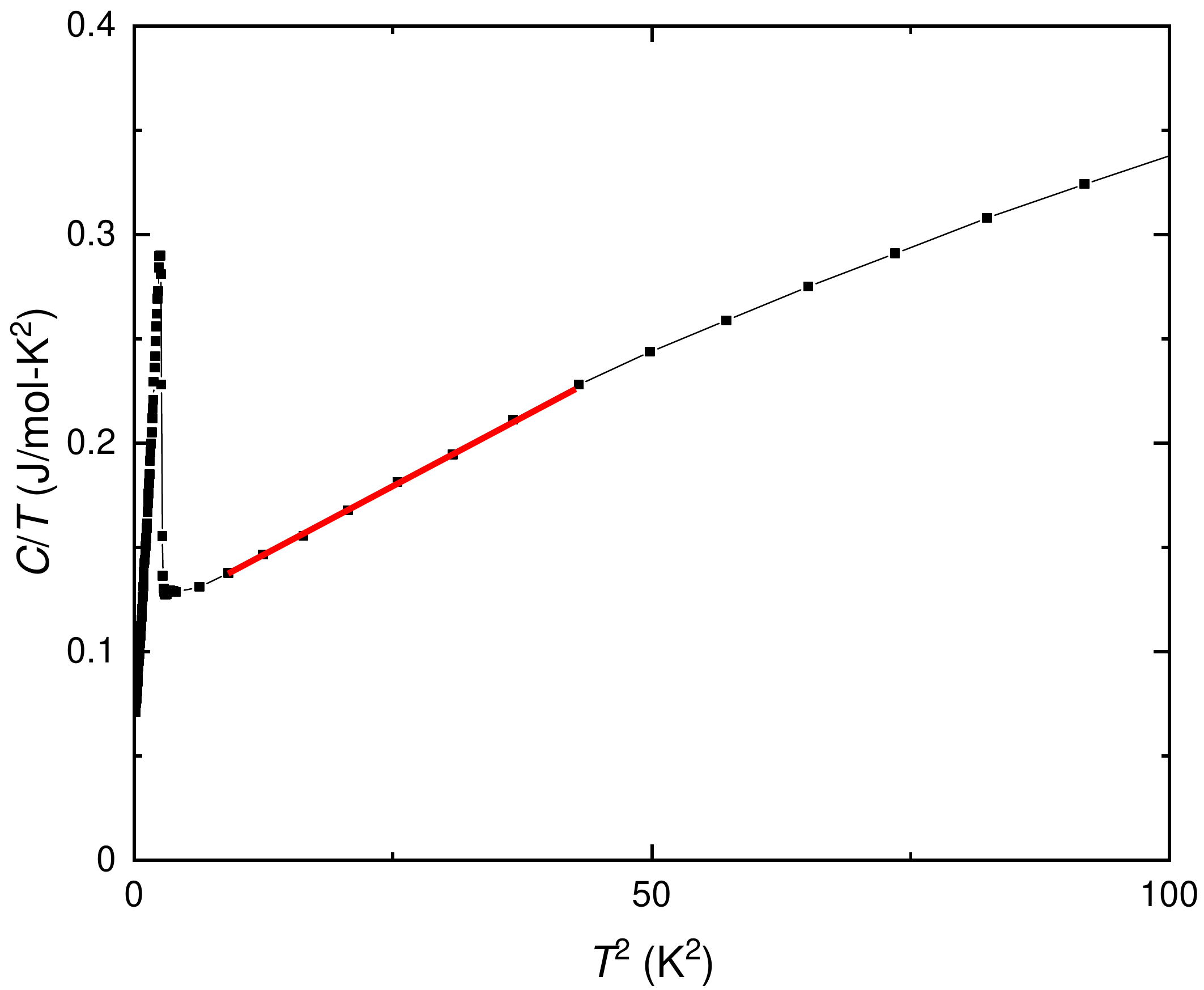}
\caption{$C/T$ data as function of $T^2$. There is a linear region above $T_c$, from which phonon contribution to the specific heat is obtained by fitting to a linear function. The red line is the fit. No magnetic order is detected above $T_c$.}
\label{Cphonon}
\end{figure}

\begin{figure}
\centering
\includegraphics[angle=0,width=100mm]{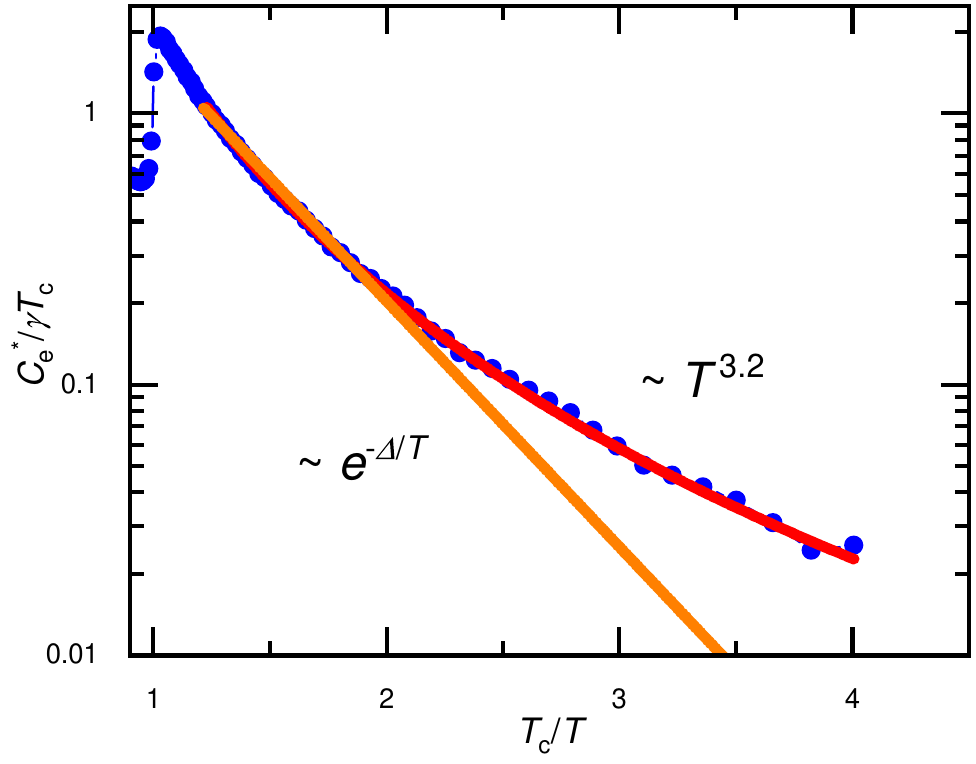}
\caption{Semilog plot of {\itshape C}$_{e}^{*}$/$\gamma T_c$ ({\itshape C}$_{e}^{*}$ is the electric contribution to specific heat minus the residue term at the zero temperature limit) as a function of {\itshape T}$_{c}$/{\itshape T}. Orange line is the fit to the BCS type of behavior. Red line is the fit to a power law with $n$ = 3.2 $\pm$ 0.1.}
\label{Cfit}
\end{figure}

\begin{figure}
\centering
\includegraphics[angle=0,width=100mm]{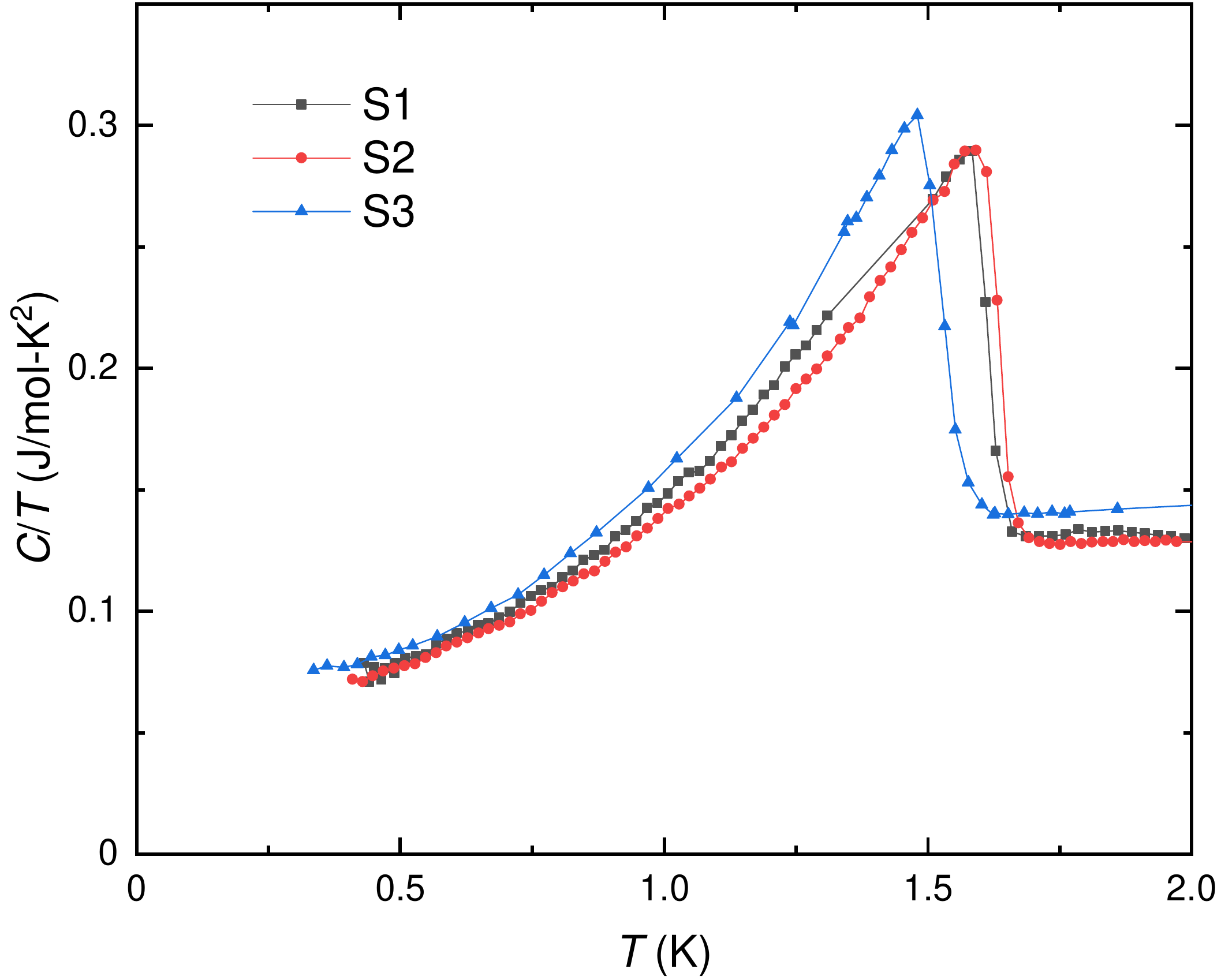}
\caption{$C/T$ data for different samples. The residue $\gamma$ in the superconducting state does not show obvious sample variation.}
\label{Cvariation}
\end{figure}

\begin{figure}
\centering
\includegraphics[angle=0,width=100mm]{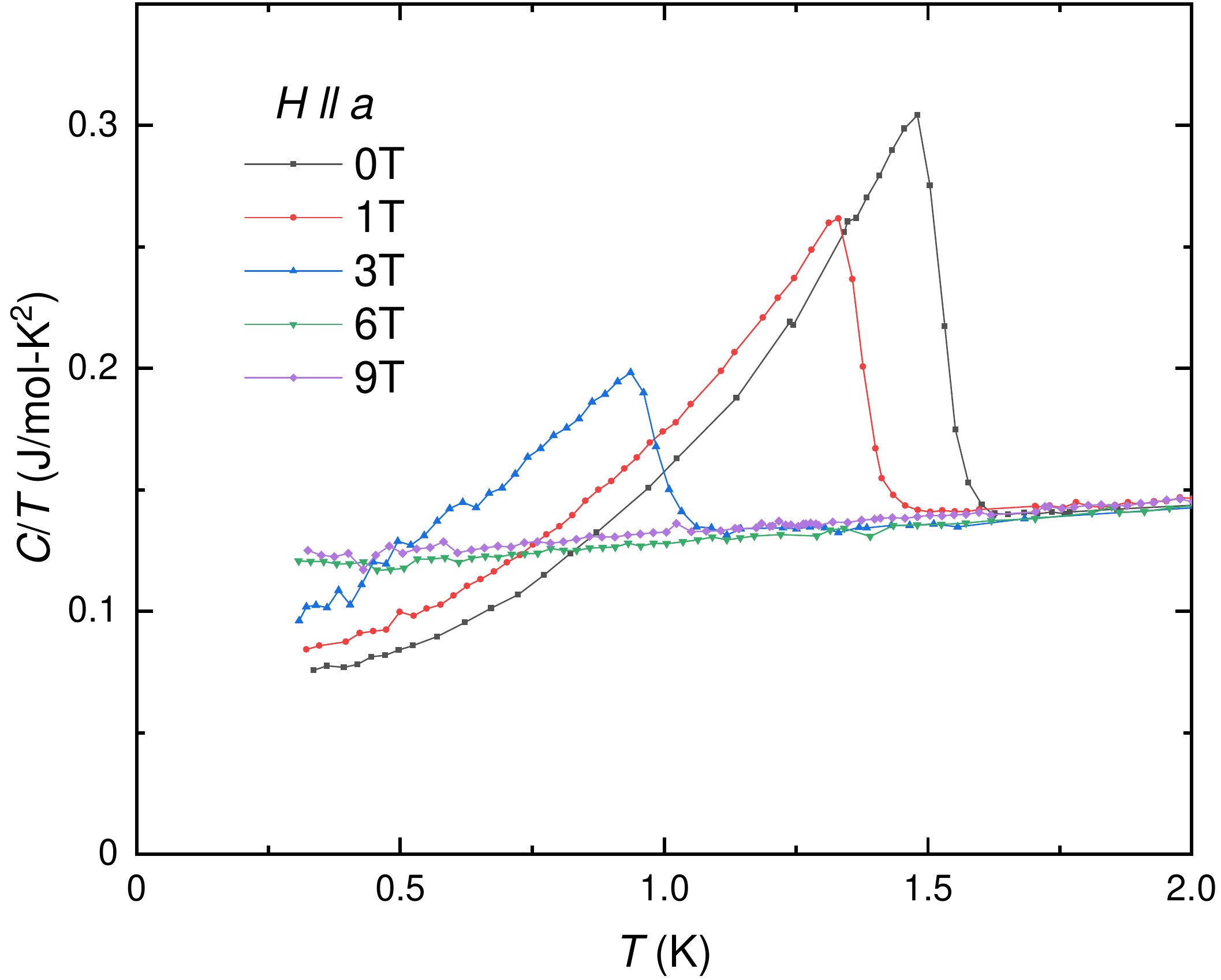}
\caption{$C/T$ data in different magnetic fields applied along $a$-axis. $H_{c2}$ is approximately 6~T in this direction. The large normal state $C/T$ is that of a heavy Fermi liquid. }
\label{Cfield}
\end{figure}

\begin{figure}
\centering
\includegraphics[angle=0,width=100mm]{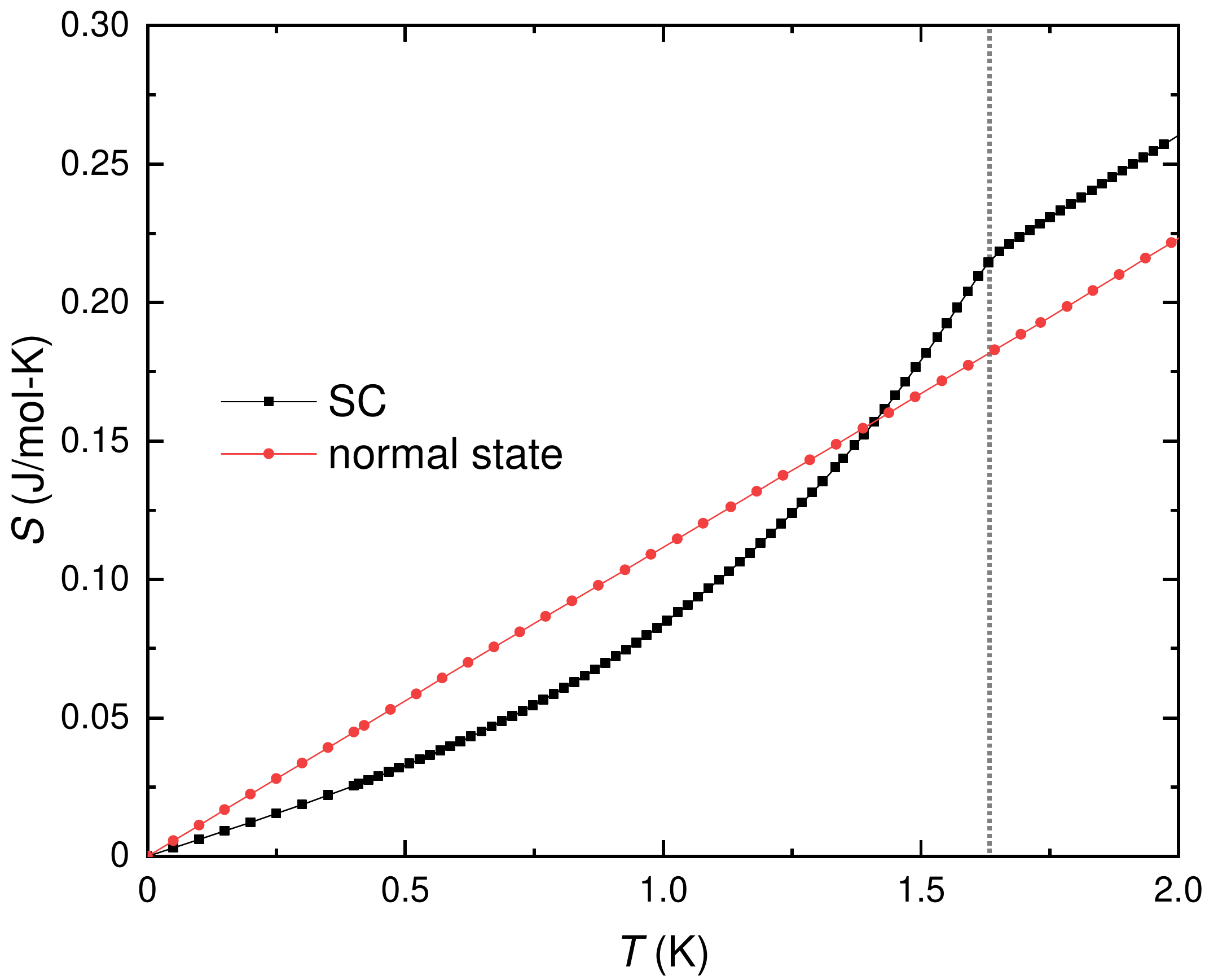}
\caption{Entropy calculated from specific heat data for superconducting and normal state. The normal state data are obtained by applying magnetic field of 7~T along the $a$-axis to suppress superconductivity.}
\label{entropy}
\end{figure}

\begin{figure}
\centering
\includegraphics[angle=0,width=100mm]{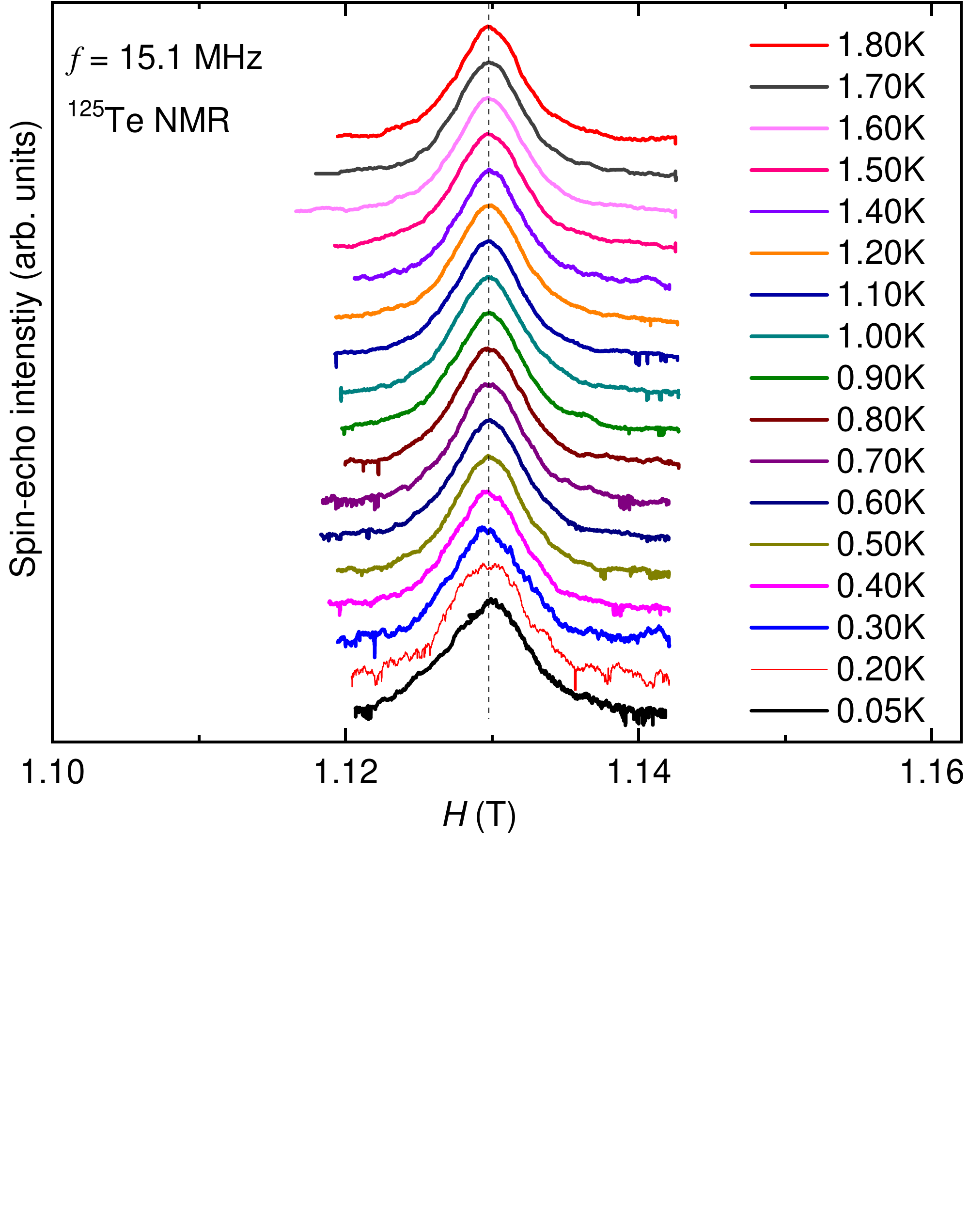}
\caption{$^{125}$Te NMR spectra in both the normal and the superconducting states of of UTe$_{2}$ at $f$ = 15.1 MHz.}
\label{NMRSM}
\end{figure}

\end{document}